\documentclass[12pt]{article}
\usepackage{amsmath,amssymb}
\allowdisplaybreaks
\usepackage{feynmp}
\usepackage{emp}
\empaddtoprelude{input graph;}
\DeclareMathOperator{\Vol}{Vol}
\DeclareMathOperator{\atan}{atan}
\makeindex
\begin{document}
\title{%
  Vegas Revisited:
  Adaptive Monte Carlo Integration Beyond Factorization}
\author{%
  Thorsten Ohl%
    \thanks{e-mail: \texttt{ohl@hep.tu-darmstadt.de}}
  {}\thanks{Supported by Bundesministerium f\"ur Bildung,
       Wissenschaft, Forschung und Technologie, Germany.}\\
  \hfil \\
  Darmstadt University of Technology  \\
  Schlo\ss gartenstr.~9 \\
  D-64289 Darmstadt \\
  Germany}
\date{%
  IKDA 98/15\\
  hep-ph/9806432\\
  June 1998}
\maketitle
\begin{abstract}
  We present a new adaptive Monte Carlo integration algorithm for
  ill-behaved integrands with non-factorizable singularities.  The
  algorithm combines Vegas with multi channel sampling and performs
  significantly better than Vegas for a large class of integrals
  appearing in physics.
\end{abstract}
%%%%%%%%%%%%%%%%%%%%%%%%%%%%%%%%%%%%%%%%%%%%%%%%%%%%%%%%%%%%%%%%%%%%%%%%
\begin{empfile}
\begin{fmffile}{\jobname pics}
\fmfset{curly_len}{2mm}
\fmfset{wiggly_len}{3mm}
%%%%%%%%%%%%%%%%%%%%%%%%%%%%%%%%%%%%%%%%%%%%%%%%%%%%%%%%%%%%%%%%%%%%%%%%
%%% (Ab)use FeynMF for drawing portable commutative diagrams
\fmfcmd{%
  style_def isomorphism expr p =
    cdraw (subpath (0, 1 - arrow_len/pixlen(p,10)) of p);
    cfill (harrow (p, 1))
  enddef;
  style_def morphism expr p =
    draw_dots (subpath (0, 1 - arrow_len/pixlen(p,10)) of p);
    cfill (harrow (p, 1))
  enddef;}
\def\fmfcd(#1,#2){%
  \begin{minipage}{#1\unitlength}%
    \vspace*{.5\baselineskip}%
    \begin{fmfgraph*}(#1,#2)%
    \fmfset{arrow_len}{3mm}%
    \fmfset{arrow_ang}{10}%
    \fmfstraight}
\def\endfmfcd{%
    \end{fmfgraph*}%
    \vspace*{.5\baselineskip}%
  \end{minipage}}
\newcommand{\fmfcdmorphism}[4]{%
  \fmf{#1,label.side=#2,label.dist=3pt,label={\small $#4$}}{#3}}
\newcommand{\fmfcdisomorph}[3][left]{%
  \fmfcdmorphism{isomorphism}{#1}{#2}{#3}}
\newcommand{\fmfcdmorph}[3][left]{%
  \fmfcdmorphism{morphism}{#1}{#2}{#3}}
\newcommand{\fmfcdeq}[1]{\fmf{double}{#1}}
\def\fmfcdsetaux[#1]#2{%
  \fmfv{decor.shape=circle,decor.size=18pt,foreground=white,
        label.dist=0,label=$#1$}{#2}}
\makeatletter
  \def\fmfcdset{\@dblarg{\fmfcdsetaux}}
\makeatother
%%%%%%%%%%%%%%%%%%%%%%%%%%%%%%%%%%%%%%%%%%%%%%%%%%%%%%%%%%%%%%%%%%%%%%%%

%%%%%%%%%%%%%%%%%%%%%%%%%%%%%%%%%%%%%%%%%%%%%%%%%%%%%%%%%%%%%%%%%%%%%%%%
\section{Introduction}

Throughout physics, it is frequently necessary to evaluate the
integral~$I(f)$ of a function~$f$ on a manifold~$M$ using a
measure~$\mu$
\begin{equation}
\label{eq:I(f)}
  I(f) = \int_M\!\textrm{d}\mu(p)\,f(p)\,.
\end{equation}
More often than not, an analytical evaluation in terms of elementary
or known special functions is impossible and we have to rely on
numerical methods for estimating~$I(f)$.  A typical example is given
by the integration of differential cross sections on a part of phase
space to obtain predictions for event rates in scattering experiments.

In more than three dimensions, standard quadrature formulae are not
practical and Monte Carlo integration is the only option.  As is well
known, $I(f)$ is estimated by
\begin{equation}
\label{eq:E(f)}
  E(f) = \left\langle \frac{f}{g} \right\rangle_g
       = \frac{1}{N} \sum_{i=1}^{N} \frac{f(p_i)}{g(p_i)}\,,
\end{equation}
where~$g$ is the probability density (with respect to the
measure~$\mu$) of the randomly distributed~$p_i$,
e.\,g.~$g(p)=1/\Vol(M)$ for uniformly distributed~$p_i$.  The error of
this estimate is given by the square root of the variance
\begin{equation}
\label{eq:V(f)}
  V(f) = \frac{1}{N-1}
       \left(\left\langle\left(\frac{f}{g}\right)^2\right\rangle_g
             - \left\langle\frac{f}{g}\right\rangle_g^2 \right)
\end{equation}
which suggests to choose a~$g$ that minimizes~$V(f)$.  If~$f$ is a
wildly fluctuating function, this optimization of~$g$ is indispensable
for obtaining a useful accuracy.  Typical causes for large
fluctuations are integrable singularities of~$f$ or~$\mu$ inside
of~$M$ or non-integrable singularities very close to~$M$.  Therefore,
we will use the term ``singularity'' for those parts
of~$M$ in which there are large fluctuations in~$f$ or~$\mu$.

Manual optimization of~$g$ is often too time consuming, in particular
if the dependence of the integral on external parameters (in the
integrand and in the boundaries) is to be studied.  Adaptive numerical
approaches are more attractive in these cases.  The problem of
optimizing~$g$ numerically has been solved for \emph{factorizable}
distributions~$g$ and measures~$\mu$ by the classic
Vegas~\cite{Lepage:1978:vegas} algorithm long ago.  Factorizable~$g$
and~$\mu$ are special, because the computational costs for
optimization rise only linearly with the number of dimensions.  In all
other cases, there is a prohibitive exponential rise of the
computational costs with the number of dimensions.

The property of factorization depends on the coordinate system, of
course. Consider, for example, the functions
\begin{subequations}
\label{eq:f1f2}
\begin{align}
  f_1(x_1,x_2) & = \frac{1}{(x_1-a_1)^2 + b_1^2} \\
  f_2(x_1,x_2) & = \frac{1}{\left(\sqrt{x_1^2+x_2^2}-a_2\right)^2 + b_2^2}
\end{align}
\end{subequations}
on~$M=(-1,1)\otimes(-1,1)$ with the
measure~$\textrm{d}\mu=\textrm{d}x_1\wedge\textrm{d}x_2$. Obviously,
$f_1$ is factorizable in Cartesian coordinates, while~$f_2$ is
factorizable in polar coordinates.  Vegas will sample either function
efficiently for arbitrary~$b_{1,2}$ in suitable coordinate systems,
but there is no coordinate system in which Vegas can sample the
sum~$f_1+f_2$ efficiently for small~$b_{1,2}$.

In this note, we present a generalization of the Vegas algorithm from
factorizable distributions to sums of factorizable distributions,
where each term may be factorizable in a \emph{different} coordinate
system.  This larger class includes most of the integrands appearing in
particle physics and empirical studies have shown a dramatic increase
of accuracy for typical integrals.  Technically, this generalization
is the combination of the Vegas algorithm with adaptive multi channel
sampling~\cite{Kleiss/Pittau:1994:multichannel}.

In section~\ref{sec:maps}, we will discuss the coordinate
transformations employed by the algorithm and in section~\ref{sec:MC},
we will describe the adaptive multi channel algorithm.  Finally, I
will discuss the performance of a first implementation of the
algorithm in section~\ref{sec:performance} and conclude.

%%%%%%%%%%%%%%%%%%%%%%%%%%%%%%%%%%%%%%%%%%%%%%%%%%%%%%%%%%%%%%%%%%%%%%%%
\section{Maps}
\label{sec:maps}

The problem of estimating~$I(f)$ can be divided naturally into two
parts: parametrization of~$M$ and sampling of the function~$f$.  While
the estimate will not depend on the parametrization, the error will.

In general, we need an atlas with more that one chart~$\phi$ to cover
the manifold~$M$.  We can ignore this technical complication in the
following, because, for the purpose of integration, we can
decompose~$M$ such that each piece is covered by a single chart.
Moreover, a single chart suffices in most cases of practical interest,
since we are at liberty to remove sets of measure zero from~$M$.  For
example, after removing a single point, the unit sphere can be covered
by a single chart.

Nevertheless, even if we are not concerned with the global properties
of~$M$ that require the use of more than one chart, the language of
differential geometry will allow us to use our geometrical intuition.
Instead of pasting together locally flat pieces, we will paste
together \emph{factorizable} pieces, which can be overlapping, because
integration is an additive operation.

For actual computations, it is convenient to use the same domain for
the charts of all manifolds.  The obvious choice for $n$-dimensional
manifolds is the open $n$-dimensional unit hypercube
\begin{equation}
  U = (0,1)^{\otimes n}\,.
\end{equation}
Sometimes, it will be instructive to view the chart as a
composition~$\phi=\psi\circ\chi$ with an irregularly
shaped~$P\in\mathbf{R}^n$ as an intermediate step
\begin{equation}
  \begin{fmfcd}(40,20)
    \fmfbottom{P,R}
    \fmftop{U}
    \fmfcdset{U}
    \fmfcdset{P}
    \fmfcdset[\mathbf{R}]{R}
    \fmfcdset{M}
    \fmfcdisomorph{U,M}{\phi}
    \fmfcdisomorph[right]{U,P}{\chi}
    \fmfcdisomorph{P,M}{\psi}
    \fmfcdmorph[right]{P,R}{f\circ\psi}
    \fmfcdmorph{M,R}{f}
    \fmfcdmorph{U,R}{f\circ\phi}
  \end{fmfcd}
\end{equation}
(in all commutative diagrams, solid arrows are reserved for
bijections and dotted arrows are used for other morphisms).
The integral~(\ref{eq:I(f)}) can now be written
\begin{equation}
  I(f) = \int_0^1\!\textrm{d}^nx\,
    \left|\frac{\partial\phi}{\partial x}\right| f(\phi(x))
\end{equation}
and it remains to sample~$|\partial\phi/\partial x|\cdot(f\circ\phi)$
on~$U$.  Below, it will be crucial that there is more than one way to
map~$U$ onto~$M$
\begin{equation}
\label{eq:pi}
  \begin{fmfcd}(40,20)
    \fmfleft{U',U}
    \fmfright{P',P}
    \fmfcdset{U}
    \fmfcdset{U'}
    \fmfcdset{P}
    \fmfcdset{P'}
    \fmfcdset{M}
    \fmfcdisomorph{U,M}{\phi}
    \fmfcdisomorph{U',M}{\phi'}
    \fmfcdisomorph[right]{P,M}{\psi}
    \fmfcdisomorph[right]{P',M}{\psi'}
    \fmfcdisomorph[right]{U,U'}{\pi_U}
    \fmfcdisomorph{P,P'}{\pi_P}
    \fmfcdisomorph{U,P}{\chi}
    \fmfcdisomorph{U',P'}{\chi'}
  \end{fmfcd}
\end{equation}
and that we are free to select the map most suitable for our purposes.

The ideal choice for~$\phi$ would be a solution of the partial
differential equation $|\partial\phi/\partial x| = 1/(f\circ\phi)$,
but this is equivalent to an analytical evaluation of~$I(f)$ and is
impossible for the cases under consideration.  A more realistic goal
is to find a~$\phi$ such that $|\partial\phi/\partial
x|\cdot(f\circ\phi)$ has factorizable singularities and is therefore
sampled well by Vegas.  This is still a non-trivial problem, however.

\begin{figure}
  \begin{center}
    \hfill\\
    \vspace*{\baselineskip}
    \begin{fmfgraph*}(30,20)
      \fmfleft{p1,p2}
      \fmfright{q1,k,q2}
      \fmflabel{$p_1$}{p1}
      \fmflabel{$p_2$}{p2}
      \fmflabel{$q_1$}{q1}
      \fmflabel{$q_2$}{q2}
      \fmflabel{$k$}{k}
      \fmf{fermion}{p1,v,p2}
      \fmf{photon}{v,vq}
      \fmf{fermion,tension=.5}{q1,vq}
      \fmf{fermion,tension=.5,label=$s_2$,label.side=left}{vq,vg}
      \fmf{fermion,tension=.5}{vg,q2}
      \fmffreeze
      \fmf{gluon}{vg,k}
      \fmfdot{v,vq,vg}
    \end{fmfgraph*}
    \qquad\qquad
    \begin{fmfgraph*}(30,20)
      \fmfleft{p1,p2}
      \fmfright{q1,k,q2} 
      \fmflabel{$p_1$}{p1}
      \fmflabel{$p_2$}{p2}
      \fmflabel{$q_1$}{q1}
      \fmflabel{$q_2$}{q2}
      \fmflabel{$k$}{k}
      \fmf{fermion}{p1,v,p2}
      \fmf{photon}{v,vq}
      \fmf{fermion,tension=.5}{q1,vg}
      \fmf{fermion,tension=.5,label=$s_1$,label.side=left}{vg,vq}
      \fmf{fermion,tension=.5}{vq,q2}
      \fmffreeze
      \fmf{gluon}{vg,k}
      \fmfdot{v,vq,vg}
    \end{fmfgraph*}
  \end{center}
  \caption{\label{fig:glue}%
    $e^+e^-\to q\bar qg$}
\end{figure}

For example, consider the phase space integration for gluon radiation
$e^+e^-\to q\bar qg$.  From the Feynman diagrams in
figure~\ref{fig:glue} it is obvious that the squared matrix element
will have singularities in the variables~$s_{1/2}=(q_{1/2}+k)^2$.
Thus, adaptive sampling using Vegas would benefit from a
parametrization using both~$s_1$ and~$s_2$ as coordinates in the
intermediate space~$P$.  Unfortunately, the invariant phase space
measure for such a parametrization involves the Gram determinant in
the form~$1/\sqrt{\Delta_4(p_1,p_2,q_1,q_2)}$, which will lead to
non-factorizable singularities at the edges of phase space.  Note that
the very elegant phase space parametrizations of the
RAMBO~\cite{Kleiss/Stirling/Ellis:1986:RAMBO} type are not useful in
this case, because there is no simple relation between the coordinates
on~$U$ and the invariants in which the squared matrix elements can
have singularities.  On the other hand, it is straightforward to find
parametrizations that factorize the dependency on~$s_1$ or~$s_2$
\emph{separately}.

Returning to the general case, consider~$N_c$ different
maps~$\phi_i:U\to M$ and probability densities~$g_i:U\to [0,\infty)$.
Then the function
\begin{equation}
\label{eq:g(p)}
  g = \sum_{i=1}^{N_c} \alpha_i
     (g_i\circ\phi_i^{-1}) \left|\frac{\partial\phi_i^{-1}}{\partial p}\right|
\end{equation}
is a probability density~$g:M\to [0,\infty)$
\begin{equation}
  \int_M\! \textrm{d}\mu(p)\, g(p) = 1\,,
\end{equation}
as long as the~$g_i$ and~$\alpha_i$ are properly normalized
\begin{equation}
\label{eq:alpha}
 \int_0^1\!g_i(x)\textrm{d}^nx = 1\,,\;\;\;
 \sum_{i=1}^{N_c} \alpha_i = 1\,,\;\;\;
   0 \le \alpha_i \le 1 \,.
\end{equation}
{}From the definition~(\ref{eq:g(p)}), we have obviously
\begin{equation}
\label{eq:I(f)MC}
  I(f) = \sum_{i=1}^{N_c} \alpha_i
      \int_M\! g_i(\phi_i^{-1}(p))
          \left|\frac{\partial\phi_i^{-1}}{\partial p}\right|
          \textrm{d}\mu(p)\,
        \frac{f(p)}{g(p)}
\end{equation}
and, after pulling back from~$M$ to~$U$
\begin{equation}
  I(f) = \sum_{i=1}^{N_c} \alpha_i
      \int_0^1\!g_i(x)\textrm{d}^nx\,
          \frac{f(\phi_i(x))}{g(\phi_i(x))}\,,
\end{equation}
we find the estimate
\begin{equation}
\label{eq:E(f)MC}
  E(f) = \sum_{i=1}^{N_c} \alpha_i
    \left\langle \frac{f\circ\phi_i}{g\circ\phi_i} \right\rangle_{g_i}\,.
\end{equation}
The factorized~$g_i$ in~(\ref{eq:I(f)MC}) and~(\ref{eq:E(f)MC}) can be
optimized using the classic Vegas algorithm~\cite{Lepage:1978:vegas}
unchanged.  However, since we have to sample with a separate adaptive
grid for each channel, a new implementation~\cite{Ohl:1998:VAMP} is
required for technical reasons.

Using the maps~$\pi_{ij}=\phi_j^{-1}\circ\phi_i:U\to U$ introduced
in~(\ref{eq:pi}), we can write the~$g\circ\phi_i:U\to[0,\infty)$
from~(\ref{eq:E(f)MC}) as
\begin{equation}
\label{eq:gophi_i}
  g\circ\phi_i
     = \left|\frac{\partial\phi_i}{\partial x}\right|^{-1}
       \left( \alpha_i g_i + 
       \sum_{\substack{j=1\\j\not=i}}^{N_c} \alpha_j (g_j\circ\pi_{ij})
          \left|\frac{\partial\pi_{ij}}{\partial x}\right| \right)\,.
\end{equation}
{}From a geometrical perspective, the maps~$\pi_{ij}$ are just the
coordinate transformations from the coordinate systems in which the
other singularities factorize into the coordinate system in which the
current singularity factorizes.

Note that the integral in~(\ref{eq:I(f)MC}) does not change, when we
use~$\phi_i:U\to M_i\supseteq M$, if we extent~$f$ from~$M$ to~$M_i$
by the definition~$f(M_i\setminus M)=0$.
This is useful, for instance, when we want to
cover~$(-1,1)\otimes(-1,1)$ by both Cartesian and polar coordinates.
This causes, however, a problem with the~$\pi_{12}$
in~(\ref{eq:gophi_i}).  In the diagram
\begin{equation}
  \begin{fmfcd}(75,15)
    \fmfbottom{d1,U1,d2,U2,d3}
    \fmftop{P1,M1,M,M2,P2}
    \fmfcdset[U]{U1}
    \fmfcdset[U]{U2}
    \fmfcdset[P_1]{P1}
    \fmfcdset[M_1]{M1}
    \fmfcdset{M}
    \fmfcdset[M_2]{M2}
    \fmfcdset[P_2]{P2}
    \fmfcdisomorph[right]{U1,M1}{\phi_1}
    \fmfcdisomorph{U1,P1}{\chi_1}
    \fmfcdisomorph{P1,M1}{\psi_1}
    \fmfcdisomorph{U2,M2}{\phi_2}
    \fmfcdisomorph[right]{U2,P2}{\chi_2}
    \fmfcdisomorph[right]{P2,M2}{\psi_2}
    \fmfcdmorph{U1,U2}{\pi_{12}}
    \fmfcdmorph[right]{M,M1}{\iota_1}
    \fmfcdmorph{M,M2}{\iota_2}
  \end{fmfcd}
\end{equation}
the injections~$\iota_{1,2}$ are not onto and since~$\pi_{12}$ is
not necessarily a bijection anymore, the
Jacobian~$\left|\partial\pi_{ij}/\partial x\right|$ may be
ill-defined.  But since~$f(M_i\setminus M)=0$, we only need 
the unique bijections~$\phi'_{1,2}$ and~$\pi'_{12}$ that make the
diagram
\begin{equation}
  \begin{fmfcd}(90,15)
    \fmfbottom{d1,U1,U1',U2',U2,d3}
    \fmftop{P1,M1,M1',M2',M2,P2}
    \fmfcdset[U]{U1}
    \fmfcdset[U_1]{U1'}
    \fmfcdset[U_2]{U2'}
    \fmfcdset[U]{U2}
    \fmfcdset[P_1]{P1}
    \fmfcdset[M_1]{M1}
    \fmfcdset[M]{M1'}
    \fmfcdset[M]{M2'}
    \fmfcdset[M_2]{M2}
    \fmfcdset[P_2]{P2}
    \fmfcdisomorph{U1',M1'}{\phi'_1}
    \fmfcdisomorph[right]{U1,M1}{\phi_1}
    \fmfcdisomorph{U1,P1}{\chi_1}
    \fmfcdisomorph{P1,M1}{\psi_1}
    \fmfcdisomorph[right]{U2',M2'}{\phi'_2}
    \fmfcdisomorph{U2,M2}{\phi_2}
    \fmfcdisomorph[right]{U2,P2}{\chi_2}
    \fmfcdisomorph[right]{P2,M2}{\psi_2}
    \fmfcdmorph{U1',U1}{\iota^U_1}
    \fmfcdisomorph{U1',U2'}{\pi'_{12}}
    \fmfcdmorph[right]{U2',U2}{\iota^U_2}
    \fmfcdmorph[right]{M1',M1}{\iota_1}
    \fmfcdeq{M1',M2'}
    \fmfcdmorph{M2',M2}{\iota_2}
  \end{fmfcd}
\end{equation}
commute.

In many applications, the dependence of an integral on external
parameters has to be studied.  Often, the~$\pi_{ij}$ will not depend
on these parameters and we can rely on Vegas to optimize the~$g_i$ for
each parameter set.  In the next section, we will show how to optimize
the~$\alpha_i$ numerically as well.

%%%%%%%%%%%%%%%%%%%%%%%%%%%%%%%%%%%%%%%%%%%%%%%%%%%%%%%%%%%%%%%%%%%%%%%%
\section{Multichannel}
\label{sec:MC}

Up to now, we have not specified the~$\alpha_i$, they are only subject
to the conditions~(\ref{eq:alpha}).  Intuitively, we expect the best
results when the~$\alpha_i$ are proportional to the contribution of their
corresponding singularities to the integral.  The option of tuning
the~$\alpha_i$ manually is not attractive if the optimal values depend
on varying external parameters.  Instead, we use a numerical
procedure~\cite{Kleiss/Pittau:1994:multichannel} for tuning
the~$\alpha_i$.

We want to minimize the variance~(\ref{eq:V(f)}) with respect to
the~$\alpha_i$. This is equivalent to minimizing
\begin{equation}
\label{eq:W(alpha)}
  W(f,\alpha) = \int_M\! g(p) \textrm{d}\mu(p)\,
       \left(\frac{f(p)}{g(p)}\right)^2
\end{equation}
with respect to~$\alpha$ with the subsidiary
condition~$\sum_i\alpha_i=1$.  After adding a Lagrange multiplier, the
stationary points of the variation are given by the solutions to the
equations
\begin{equation}
\label{eq:PDG(W)}
  \forall i: W_i(f,\alpha) = W(f,\alpha)
\end{equation}
where
\begin{equation}
  W_i(f,\alpha)
    = -\frac{\partial}{\partial\alpha_i} W(f,\alpha)
    = \int_0^1\!g_i(x)\textrm{d}^nx\,
          \left(\frac{f(\phi_i(x))}{g(\phi_i(x))}\right)^2
\end{equation}
and
\begin{equation}
   W(f,\alpha) = \sum_{i=1}^{N_c} \alpha_i W_i(f,\alpha)\,.
\end{equation}
It can easily be shown~\cite{Kleiss/Pittau:1994:multichannel} that the
stationary points~(\ref{eq:PDG(W)}) correspond to local minima.
If we use
\begin{equation}
   N_i = \alpha_i N
\end{equation}
to distribute~$N$ sampling points among the channels,
the~$W_i(f,\alpha)$ are just the contributions from channel~$i$ to the
total variance.  Thus we recover the familiar result from
stratified sampling, that the overall variance is minimized by
spreading the variance evenly among channels.

The~$W_i(f,\alpha)$ can be estimated with very little extra effort
while sampling~$I(f)$ (cf.~\ref{eq:E(f)MC})
\begin{equation}
\label{eq:Vi(alpha)}
  V_i(f,\alpha) =
    \left\langle \left(\frac{f\circ\phi_i}{g\circ\phi_i}\right)^2
       \right\rangle_{g_i}\,.
\end{equation}
Note that the factor of~$g_i/g$ from the corresponding formula
in~\cite{Kleiss/Pittau:1994:multichannel} is absent
from~(\ref{eq:Vi(alpha)}), because we are already sampling with the
weight~$g_i$ in each channel separately.

The equations~(\ref{eq:PDG(W)}) are a fixed point of the prescription
\begin{equation}
\label{eq:update}
  \alpha_i \mapsto \alpha_i'
      = \frac{\alpha_i \left(V_i(f,\alpha)\right)^\beta}
             {\sum_i\alpha_i \left(V_i(f,\alpha)\right)^\beta},
  \;\;\;(\beta>0)
\end{equation}
for updating the weights~$\alpha_i$.  There is no guarantee that this
fixed point will be reached from a particular starting value, such
as~$\alpha_i=1/N_c$, through successive applications
of~(\ref{eq:update}).  Nevertheless, it is clear
that~(\ref{eq:update}) will concentrate on the channels with large
contributions to the variance, as suggested by stratified
sampling. Furthermore, empirical studies show that~(\ref{eq:update})
is successful in practical applications.
The value~$\beta=1/2$ has been proposed
in~\cite{Kleiss/Pittau:1994:multichannel}, but it can be beneficial in
some cases to use smaller values like~$\beta=1/4$ to dampen
statistical fluctuations.

%%%%%%%%%%%%%%%%%%%%%%%%%%%%%%%%%%%%%%%%%%%%%%%%%%%%%%%%%%%%%%%%%%%%%%%%
\section{Performance}
\label{sec:performance}

Both the implementation and the practical use of the algorithm
proposed in this note are more involved than the
application of the original Vegas algorithm.  Therefore it is
necessary to investigate whether the additional effort pays off in
terms of better performance.

A test version of an implementation of this algorithm, ``VAMP'', in
Fortran~\cite{Fortran95} has been used for empirical studies.  This
implementation features other improvements over ``Vegas
Classic''---most notably system independent and portable support for
parallel processing and support for unweighted event generation---and
will be published when the documentation~\cite{Ohl:1998:VAMP} is
finalized.  The preliminary version is available from the author upon
request.

%%%%%%%%%%%%%%%%%%%%%%%%%%%%%%%%%%%%%%%%%%%%%%%%%%%%%%%%%%%%%%%%%%%%%%%%
\subsection{Costs}

There are two main sources of additional computational costs: at each
sampling point the function~$g\circ\phi_i$ has be evaluated, which
requires the computation of the~$N_c-1$ maps~$\pi_{ij}$ together with
their Jacobians and of the~$N_c-1$ probability distributions~$g_i$ of
the other Vegas grids (cf.~(\ref{eq:gophi_i})).

The retrieval of the current~$g_i$s requires a bisection search in
each dimension, i.e.~a total of~$O((N_c-1)\cdot n_{\text{dim}}\cdot
\log_2 (n_{\text{div}}))$ executions of the inner loop of the search.
For simple integrands, this can indeed be a few times more costly than
the evaluation of the integrand itself.

The computation of the~$\pi_{ij}$ can be costly as well.  However,
unlike the~$g_i$, this computation can usually be tuned manually.
This can be worth the effort if many estimations of similar integrals
are to be performed.  Empirically, straightforward implementations of
the~$\pi_{ij}$ add costs of the same order as the evaluation of
the~$g_i$.

Finally, additional iterations are needed for adapting the
weights~$\alpha_i$ of the multi channel algorithm described
in~(\ref{sec:MC}).  Their cost is negligible, however, because
they are usually performed with far fewer sampling points than the
final iterations.

%%%%%%%%%%%%%%%%%%%%%%%%%%%%%%%%%%%%%%%%%%%%%%%%%%%%%%%%%%%%%%%%%%%%%%%%
\subsection{Gains}

Even in cases in which the evaluation of~$g_i$ increases computation
costs by a whole order of magnitude, any reduction of the error by
more than a factor of~4 will make the multi channel algorithm
economical.  In fact, it is easy to construct examples in which the
error will be reduced by more than two orders of magnitude.  The
function
\begin{multline}
\label{eq:f(x)}
  f(x) = \frac{b}{144\atan(1/2b)}
      \biggl( \frac{3\pi\Theta(r_3<1)}{r_3^2((r_3-1/2)^2+b^2)}
             + \frac{2\pi\Theta(r_2<1,|x_3|<1)}{r_2((r_2-1/2)^2+b^2)} \\
             + \frac{\Theta(-1<x_1,x_2,x_3<1)}{x_1^2+b^2} \biggl)\,,
\end{multline}
with~$r_2=\sqrt{x_1^2+x_2^2}$ and~$r_3=\sqrt{x_1^2+x_2^2+x_3^2}$, is
constructed such that it can easily be normalized
\begin{equation}
  \int_{-1}^{1}\!\textrm{d}^3x\,f(x) = 1
\end{equation}
and allows a check of the result.  The three terms factorize in
spherical, cylindrical and Cartesian coordinates, respectively,
suggesting a three channel approach.  After five steps of weight
optimization consisting four iterations of $10^5$ samples, we have
performed three iterations of $10^6$ samples with the VAMP multi
channel algorithm.  Empirically, we found that we can perform four
iterations of $5\cdot10^5$ samples and three
iterations of $5\cdot10^6$ samples with the class Vegas algorithm
during the same time period.  Since the functional form of~$f$ is
almost as simple as the coordinate transformation, the fivefold
increase of computational cost is hardly surprising.

\begin{figure}
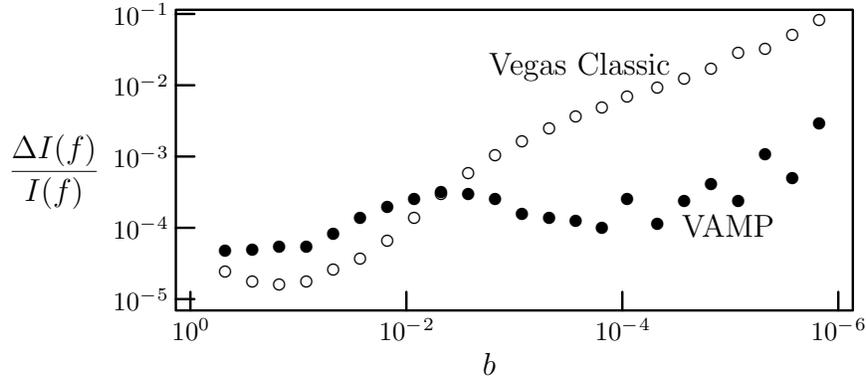

  \begin{center}
    \begin{empgraph}(90,40)
      vardef Formgen_ (expr q) = Formsci_ (q) enddef;
      pickup pencircle scaled 1pt;
      setcoords (-log, log);
      gdraw "vamp.data" plot btex $\bullet$ etex;
      glabel.lrt (btex VAMP etex, 17);
      gdraw "vegas.data" plot btex $\circ$ etex;
      glabel.ulft (btex Vegas Classic etex, 17);
      autogrid (itick.bot, itick.lft);
      glabel.lft (btex $\displaystyle\frac{\Delta I(f)}{I(f)}$ etex, OUT);
      glabel.bot (btex $b$ etex, OUT);
    \end{empgraph}
  \end{center}
  \caption{\label{fig:bench}%
    Comparison of the sampling error for the integral of~$f$
    in~(\ref{eq:f(x)}) as a function of the width parameter~$b$ for
    the two algorithms at comparable computational costs.}
\end{figure}

In figure~\ref{fig:bench}, we compare the error estimates derived by
the classic Vegas algorithm and by the three channel VAMP algorithm.
As one would expect, the multi channel algorithm does not offer any
substantial advantages for smooth functions (i.\,e.~$b>0.01$).
Instead, it is penalized by the higher computational costs.  On the
other hand, the accuracy of the classic Vegas algorithm deteriorates
like a power with smaller values of~$b$.  At the same time, the
multi channel algorithm can adapt itself to the steeper functions,
leading to a much slower loss of precision.

The function~$f$ in~(\ref{eq:f(x)}) has been constructed as a showcase
for the multi channel algorithm, of course.  Nevertheless, more
complicated realistic examples from particle physics appear to gain
about an order of magnitude in accuracy.  Furthermore, the new
algorithm allows \emph{unweighted} event generation.  This is hardly
ever possible with the original Vegas implementation, because the
remaining fluctuations typically reduce the average weight to very
small numbers.

%%%%%%%%%%%%%%%%%%%%%%%%%%%%%%%%%%%%%%%%%%%%%%%%%%%%%%%%%%%%%%%%%%%%%%%%
\subsection{A Cheaper Alternative}

There is an alternative approach that avoids the evaluation of
the~$g_i$s, sacrificing flexibility.  Fixing the~$g_i$ at unity, we
have for~$\tilde g:M\to [0,\infty)$
\begin{equation}
\label{eq:tildeg(p)}
  \tilde g = \sum_{i=1}^{N_c} \alpha_i
     \left|\frac{\partial\phi_i^{-1}}{\partial p}\right|
\end{equation}
and the integral becomes
\begin{equation}
  I(f) = \sum_{i=1}^{N_c} \alpha_i
      \int_M\! \left|\frac{\partial\phi_i^{-1}}{\partial p}\right|
          \textrm{d}\mu(p)\, \frac{f(p)}{\tilde g(p)}
       = \sum_{i=1}^{N_c} \alpha_i \int_0^1\!\textrm{d}^nx\,
          \frac{f(\phi_i(x))}{\tilde g(\phi_i(x))}\,.
\end{equation}
Vegas can now be used to perform adaptive integrations of
\begin{equation}
  I_i(f) = \int_0^1\!\textrm{d}^nx\,
          \frac{f(\phi_i(x))}{\tilde g(\phi_i(x))}
\end{equation}
individually.  In some cases it is possible to construct a set
of~$\phi_i$ such that~$I_i(f)$ can estimated efficiently.
The optimization of the weights~$\alpha_i$ can again be effected by
the multi channel algorithm described in~(\ref{sec:MC}).

The disadvantage of this approach is that the optimal~$\phi_i$ will
depend sensitively on external parameters and the integration limits.
In the approach based on the~$g$ in~(\ref{eq:g(p)}) Vegas can take
care of the integration limits automatically.

%%%%%%%%%%%%%%%%%%%%%%%%%%%%%%%%%%%%%%%%%%%%%%%%%%%%%%%%%%%%%%%%%%%%%%%%
\section{Conclusions}
\label{sec:conclusions}

We have presented an algorithm for adaptive Monte Carlo integration of
functions with non-factorizable singularities.  The algorithm shows
a significantly better performance for many ill-behaved integrals than
Vegas.

The applications of this algorithm are not restricted to particle
physics, but a particularly attractive application is provided by
automated tools for the calculation of scattering cross sections.
While these tools can currently calculate differential cross sections
without manual intervention, the phase space integrations still
require hand tuning of mappings for importance sampling for each
parameter set.  The present algorithm can overcome this problem, since
it requires to solve the geometrical problem of calculating the
maps~$\pi_{ij}$ in~(\ref{eq:gophi_i}) for all possible invariants only
\emph{once}.  The selection and optimization of the channels can then
be performed algorithmically.

The application of the algorithms presented here to quasi Monte Carlo
integration forms an interesting subject for future research.  Other
options include maps~$\phi_i$ depending on external parameters, which
can be optimized as well.  A simple example are rotations, which can
align the coordinate systems with the singularities, using
correlation matrices~\cite{Ohl:1998:VAMP}.

%%%%%%%%%%%%%%%%%%%%%%%%%%%%%%%%%%%%%%%%%%%%%%%%%%%%%%%%%%%%%%%%%%%%%%%%
%%% \bibliography{jpsi}

\begin{thebibliography}{10}
\bibitem{Lepage:1978:vegas}
  G.~P.~Lepage, J.~Comp.\ Phys.\ \textbf{27}, 192 (1978);
  G.~P.~Lepage, Cornell Preprint, CLNS-80/447, March 1980.
\bibitem{Kleiss/Pittau:1994:multichannel}
  R.~Kleiss, R.~Pittau,
  Comp.\ Phys.\ Comm.\ \textbf{83}, 141 (1994).
\bibitem{Kleiss/Stirling/Ellis:1986:RAMBO}
  R. Kleiss, W. J. Stirling, S. D. Ellis,
  Comp.\ Phys.\ Comm.\ \textbf{40}, 359 (1986);
  R. Kleiss, W. J. Stirling,
  Nucl.\ Phys.\ \textbf{B385}, 413 (1992).
\bibitem{Ohl:1998:VAMP}
  T.~Ohl,
  \textit{\texttt{VAMP}, Version 1.0: Vegas AMPlified:
    Anisotropy, Multi-channel sampling and Parallelization},
  Preprint, Darmstadt University of Technology, 1998 (in preparation).
\bibitem{Fortran95}
   International Standards Organization,
   \textit{ISO/IEC 1539:1997, Information technology --- Programming
     Languages --- Fortran,}
   Geneva, 1997.
\end{thebibliography}

%%%%%%%%%%%%%%%%%%%%%%%%%%%%%%%%%%%%%%%%%%%%%%%%%%%%%%%%%%%%%%%%%%%%%%%%
\end{fmffile}
\end{empfile}
\end{document}